\documentclass[nofootinbib,a4paper,pra,reprint,twocolumn,preprintnumbers,amssymb,10pt, superscriptaddress]{revtex4-2}

\usepackage{amsmath}
\usepackage{verbatim}
\usepackage{graphicx}
 \graphicspath{ {figs/}} 
 \usepackage{color}
\usepackage{tikz}
\usepackage{braket, enumitem}
\usepackage[normalem]{ulem}
\usepackage{upgreek}

\newcommand{\tc}[1]{\textcolor{black}{#1}}
\newcommand{\jz}[1]{\textcolor{black}{#1}}

\usepackage[colorlinks=true,citecolor=blue,linkcolor=blue,urlcolor=cyan]{hyperref}

\begin{document}

\title{
Stability of the symmetry-protected topological phase and Ising transitions in a disordered U(1) quantum link model on a ladder
}

\author{Mykhailo V. Rakov}
\affiliation{Instytut Fizyki Teoretycznej, Uniwersytet Jagiello\'nski, ulica Łojasiewicza 11, 30-348 Krak\'ow, Poland}

\author{Luca Tagliacozzo}
\affiliation{Instituto de Fisica Fundamental IFF-CSIC, Calle Serrano 113b, Madrid 28006, Spain}

\author{Maciej Lewenstein}
\affiliation{Institut de Ciencies Fotoniques (ICFO), Av. Karl Friedrich Gauss 3, 08860 Barcelona, Spain}

\author{Jakub Zakrzewski}
\affiliation{Instytut Fizyki Teoretycznej, Uniwersytet Jagiello\'nski, ulica Łojasiewicza 11, 30-348 Krak\'ow, Poland}
\affiliation{Mark Kac Complex Systems Research Center, Uniwersytet Jagiello{\'n}ski, 30-348  Krak{\'o}w, Poland}

\author{Titas Chanda}
\affiliation{Department of Physics, Indian Institute of Technology Madras, Chennai 600036, India}
\affiliation{Center for Quantum Information, Communication and Computation (CQuICC),
Indian Institute of Technology Madras, Chennai 600036, India}

\begin{abstract}
We revisit the U(1) quantum link model in a ladder geometry, finding, by finite-size scaling, that the critical exponent $\nu=1$ and the central charge $c=1/2$ are consistent with the Ising universality class for all phase transitions observed. A blind application of the Harris criterion would suggest that this criticality is lost in the presence of the disorder. It turns out not to be the case. For the disorder affecting ladder's rung hoppings only, we have found that the transitions survive disappearing only for quite strong disorder. The disorder in the ladder's legs destroys the nonzero mass phase criticality, while the symmetry-protected topological phase for zero mass survives a small disorder. The observed robustness against disorder is explained qualitatively using field-theoretic arguments.
\end{abstract}

\maketitle

\section{Introduction}

{Simulation of lattice gauge theories (LGTs) in cold-atom schemes has become a subject
of intensive recent studies \cite{Zohar2012, Tagliacozzo2013, Tagliacozzo2013a, Wiese2013, Zohar2015, Dalmonte2016, Banuls2020, Aidelsburger2021} once it was realized that some open questions and phenomena inaccessible in the high energy domain may be simulated effectively in cold atom set-ups. Typically, these proposals rely on the microscopic realization of gauge theories, resulting in a local gauge invariance. Several platforms have been proposed, most notably those in Rydberg atom arrays that allow for great flexibility in realizing different effective interactions~\cite{Tagliacozzo2013, Tagliacozzo2013a, Surace2020, Cheng2023, Homeier2023, Xu2024, Meurice2024, Halimeh25}. In effect, successful experimental demonstrations of one-dimensional gauge systems have been achieved~\cite{Martinez2016, Schweizer2019, Yang2020, Nguyen2022, Riechert2022, Chisholm2022, Bauer2023} accompanied by several simulations typically performed with tensor network (TN) techniques~\cite{Schollwoeck2011, Orus2014, Ran2020}. 
Recently, TN methods, free from the limitations of Monte Carlo approaches,
have been a tour de force in studies of low-dimensional lattice gauge theory calculations~\cite{banuls_jhep_2013, buyens_prl_2014, kuhn_pra_2014, banuls_prd_2015, buyens_prd_2016, pichler_prx_2016, banuls_prl_2017, buyens_prx_2016, chanda20, Magnifico2020, Felser2020, chanda2021, Montangero2021, Magnifico2021, chanda2024, Cataldi2024, Magnifico2024}.}

\begin{figure}[t!]
    \centering
    \includegraphics[width=\linewidth]{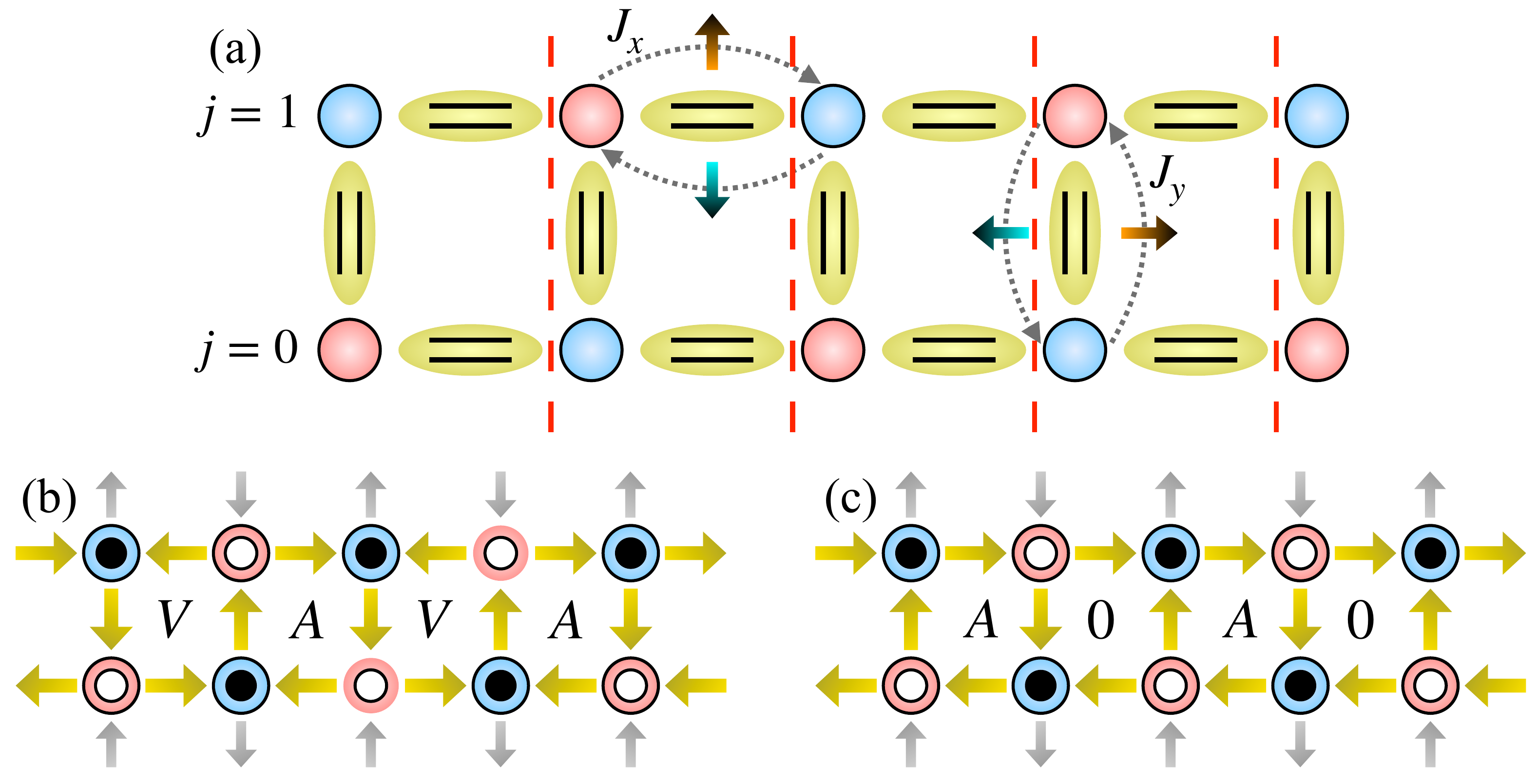}
    \caption{(Color online.) 
    (a) Schematic depiction of the ladder system described by Eq.~\eqref{eq:hamil}. Red and blue circles denote even and odd staggered fermion sites with masses $+m$ and $-m$, respectively. The bonds between sites host truncated U(1) gauge fields, represented by spin-1/2 degrees of freedom. Nearest-neighbor fermion tunneling is mediated by parallel transporters associated with the gauge fields on the intervening links, ensuring that Gauss's law is preserved.  
    (b)–(c) Two distinct phases of the system in the disorder-free case, i.e., $J_{yi} = J_y$ for all $i$, in the large-mass limit $|m| \gg J_y$. For $m > 0$, fermions occupy the odd sites (blue circles), while the even sites (red circles) remain empty, corresponding to a filled Dirac sea.  The horizontal red dashed lines indicate the partitions where we divide the system into two parts to calculate the entanglement entropy.
    (b) For smaller $J_y$, the spin configuration forms vortices and antivortices on alternating plaquettes, defining the VA phase.  
    (c) For larger $J_y$, vortices (or antivortices) appear on alternating plaquettes, while the intermediate plaquettes carry zero vorticity, defining the V0 phase.  
    In (b) and (c), grey arrows indicate the fixed virtual electric fields located just outside the ladder used to impose the boundary conditions. In our convention, $\hat{S}^z\ket{\uparrow} = +\frac{1}{2}\ket{\uparrow}$ and $\hat{S}^z\ket{\downarrow} = -\frac{1}{2}\ket{\downarrow}$ for vertical links, while $\hat{S}^z\ket{\rightarrow} = +\frac{1}{2}\ket{\rightarrow}$ and $\hat{S}^z\ket{\leftarrow} = -\frac{1}{2}\ket{\leftarrow}$ for horizontal links.
    }
    \label{fig:schem}
\end{figure}
Among various attempts, a quantum link model \cite{Wiese2013} has served over the years as a useful tool to overcome the infinite Hilbert space of the electric field and replace it with an appropriate finite-dimensional space. This allows one to go beyond a one-dimensional model and consider, e.g., two-leg ladders as the simplest (albeit pathological to some degree) two-dimensional situations. A link model in such a configuration was considered a few years ago \cite{Cardarelli2017, CardarelliThesis, Cardarelli2020}. It was found that for the zero mass, the model supports a symmetry-protected topological phase surrounded, either for a finite mass or for vastly different tunnelings between ru\jz{n}gs and legs of the ladder by vortex phases. In our work, we revisit this model (Section~\ref{sec:model}), pointing out, by detailed TN-based calculations, that the resulting transitions are of the Ising class (Section~\ref{sec:phase-diagram}). Then we consider (Section~\ref{sec:disorder}) the effects resulting from an imposed disorder. Here, let us note in passing that it is not common to study disorder-induced phenomena in the context of LGT, as they are not expected in high-energy physics, contrary to the condensed-matter domain, where LGT applications are also important. For example, many body localization in LGT systems is quite unusual \cite{Giudici20, Sau24}. Here we consider the fate of the ground state phases beyond one dimension in the presence of disorder in the LGT link model, the subject not addressed before as far as we know. \tc{\jz{Apart from pure curiosity, our study is motivated by the fact that in the experiments such a disorder appears quite naturally.} The atoms are typically held by optical tweezers in the relevant experimental setups (as, e.g., in~\cite{Halimeh25}). These optical tweezers have finite accuracy and therefore small disorder is inherent in the physics of the system.}.

We consider first in detail the case when the disorder is
imposed on the rungs tunnelings (Section~\ref{sec:disorder}) and reveal that a small disorder does not affect the symmetry-protected topological phase - such a disorder does not break the symmetry of the problem. One needs sufficiently strong disorder to destroy the SPT phase as well as mix different vortex phases present for a finite mass. We find that for small disorder, all phase transitions are in the Ising universality class by finding their half-integer central charge. The situation is markedly different when the disorder is introduced to tunnelings along the legs of the ladder (Section~\ref{sec:disx}). Here, any amount of disorder in the tunnelings for a nonzero mass kills the phase transition between different vortex states. As discussed, this is to be expected based on the quasi one-dimensional setting. Surprisingly, for a vanishing mass and a small disorder in the legs tunnelings, we still observe the existence of the SPT phase. The latter, on the other hand, trivially disappears if we add any diagonal disorder, which is equivalent to making the mass random. Finally we conclude our studies in Section~\ref{sec:conclusion}.

\section{The System}
\label{sec:model}

We consider a truncated U(1) lattice gauge theory (LGT)~\cite{Wilson1974, Kogut1975, Kogut1983, Wiese2013, Zohar2015, Banuls2020, Aidelsburger2021} on a two-leg ladder, where the gauge fields are restricted to two levels, effectively representing spin-1/2 degrees of freedom on the links connecting neighboring sites. The Hamiltonian of the corresponding quantum link model (QLM) on a two-leg ladder~\cite{Cardarelli2017, CardarelliThesis, Cardarelli2020} of length $L$ is given by
\begin{align}
    H =\ & m \sum_{i, j} (-1)^{i+j} \psi_{(i,j)}^{\dag} \psi_{(i,j)}  \nonumber \\
	&- \sum_{i, j} J_{xij}\left[\psi_{(i,j)}^{\dag} S^{+}_{(i,j),(i+1,j)} \psi_{(i+1,j)} + {\rm h.c.} \right] \nonumber \\
	&- \sum_{i} J_{yi} \left[\psi_{(i,0)}^{\dag} S^{+}_{(i,0),(i,1)} \psi_{(i,1)} + {\rm h.c.} \right],
    \label{eq:hamil}
\end{align}
where $\psi_{(i,j)}^{\dag}$ and $\psi_{(i,j)}$ are the (staggered) fermion creation and annihilation operators at rung $i = 0, 1, \ldots, L-1$ and leg $j = 0, 1$. Here, $J_{xij}$ denotes the nearest-neighbor (NN) tunneling strength along the legs of the ladder, $J_{yi}$ is the rung-dependent NN tunneling amplitude, and $m$ is the staggered mass, which is positive for even $i+j$ and negative for odd $i+j$.
The operators $S^{\pm}_{(i,j),(i+1,j)}$ and $S^{\pm}_{(i,j),(i,j+1)}$ are ladder operators associated with the spin degrees of freedom residing on the links between sites $(i,j)$ and $(i+1,j)$, and between $(i,j)$ and $(i,j+1)$, respectively. These operators act as the parallel transporters of the U(1) gauge field in the LGT. See Fig.~\ref{fig:schem}(a) for a schematic illustration of the system.

In most of the work (except Section~\ref{sec:disx}), we shall consider a constant tunneling rate along the legs, $J_{xij}=J_x$. Therefore,
for convenience, we fix this hopping to $J_x = 1$, thereby setting the unit of energy throughout our discussion. The rung-dependent hopping strengths $J_{yi}$ are drawn from a uniform random distribution over the interval $[\bar{J}_y - \delta J_y,\, \bar{J}_y + \delta J_y]$, where $\delta J_y$ is the disorder strength.

Due to the U(1) gauge invariance, the local Gauss law generators $G_{(i,j)}$, defined as
\begin{align}
    G_{(i, j)} = \Big(S^z_{(i, j), (i+1, j)} - S^z_{(i-1, j), (i, j)} + S^z_{(i, j), (i, j+1)}  \nonumber \\
    - S^z_{(i, j-1), (i, j)} \Big) - \left(\psi_{(i,j)}^{\dag} \psi_{(i,j)} - \frac{1 - (-1)^{i+j}}{2} \right),
\end{align}
commute with the Hamiltonian for all $(i, j)$. Here, $S^z = \frac{1}{2}\sigma^z$ denotes the standard spin-1/2 operator, with $\sigma^z$ being the Pauli matrix. These operators represent the truncated electric field on each link.
We restrict our analysis to the physical subspace of states that satisfy Gauss's law: $G_{(i,j)} \ket{\psi} = 0$ for all $(i, j)$. We impose open boundary conditions (OBC) along the $x$ direction. To ensure that Gauss' law is satisfied at all sites, the (virtual) electric fields just outside the ladder, along both the $x$- and $y$-directions, must be fixed appropriately. These boundary values must follow a staggered pattern in both directions (see Fig.~\ref{fig:schem}).

\begin{figure}[t]
	\includegraphics[width=\linewidth]{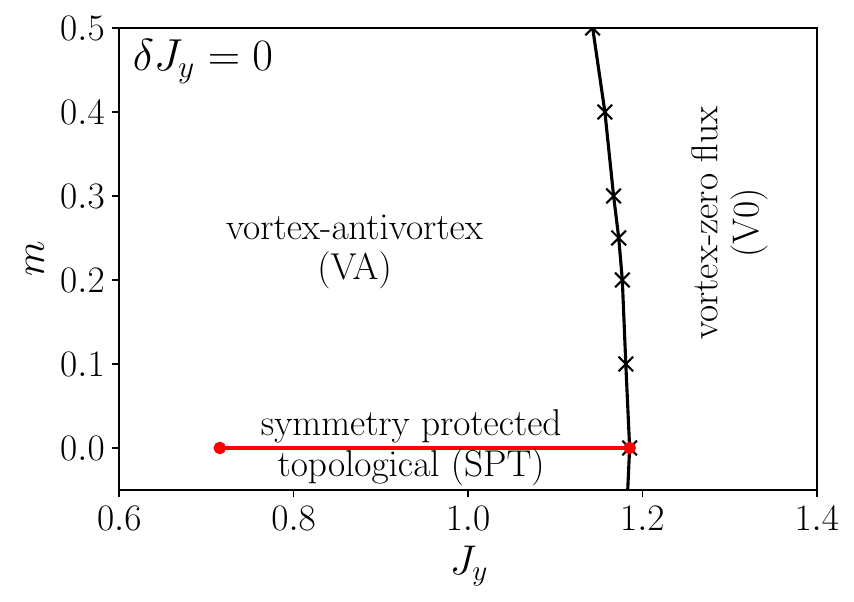}	
	\caption{(Color online.) Phase diagram of the QLM on a two-leg ladder (Eq.~\eqref{eq:hamil}) in the disorder-free limit ($\delta J_y = 0$). The diagram is symmetric under a sign flip of the fermion mass, $m \leftrightarrow -m$, so only the region $m \ge 0$ is shown. For $m > 0$, the system exhibits two quantum phases: the vortex–antivortex (VA) phase at smaller $J_y$ and the vortex–zero-flux (V0) phase at larger $J_y$. In the thermodynamic limit, the VA phase is characterized by vanishing leg magnetization $S_L$. The two phases are separated by a symmetry-protected topological (SPT) phase at $m = 0$, where both $S_L$ and rung magnetization $S_R$ as well as long-range parity order parameter $O_P$ vanish.
    }
	\label{fig:phase-no-disorder}
\end{figure}

The phase diagram of the system in the clean, disorder-free case, i.e., $J_{yi} = J_y$ for all $i$, has been studied extensively in \cite{Cardarelli2017, Cardarelli2020}. In the large-mass limit $|m| \gg J_y$, fermions occupy alternate sites, determined by the sign of $m$, to fill the Dirac sea, and Gauss's law is satisfied for two distinct configurations of the gauge fields, giving rise to two different phases. 
For smaller values of $J_y$, the spin configuration forms vortices and antivortices on alternating plaquettes (Fig.~\ref{fig:schem}(b)), referred to as the vortex-antivortex (VA) phase. In contrast, for larger $J_y$, vortices (or antivortices) appear on alternating plaquettes, while the plaquettes in between carry zero vorticity (Fig.~\ref{fig:schem}(b)). This latter phase is referred to as the vortex-zero flux (V0) phase.

To distinguish between these two phases, one can introduce the integer-valued `onsite' magnetization on each leg as  
\begin{equation}
	\tilde{S}_{(i,j)}^z = (-1)^i \left[ S_{(i-1,j),(i,j)}^z + S_{(i,j),(i+1,j)}^z \right],
    \label{eq:onsite-mag}
\end{equation}
and define the leg magnetization as  
\begin{equation}
    \tc{S_L = \frac{1}{2} \left[\frac{1}{L} \sum_i \left| \langle \tilde{S}_{(i,0)}^z \rangle \right| + \frac{1}{L} \sum_i \left| \langle \tilde{S}_{(i,1)}^z \rangle \right|\right].}
    \label{eq:SL}
\end{equation}
\tc{The two terms in brackets are equal for non-disordered case and at presence of rung disorder (by symmetry of the ladder) but unequal at leg disorder.} In the VA phase, $S_L$ vanishes in the thermodynamic limit, whereas it remains finite in the V0 phase.  
We can also define the rung magnetization as  
\begin{equation}
	S_R = \frac{1}{N} \sum_i \left| \langle S_{\tc{(i,0),(i,1)}}^z \rangle \right|,
    \label{eq:SR}
\end{equation}
which exhibits complementary behavior to $S_L$, i.e.,  $S_R$ is finite in the VA phase, but vanishes in the V0 phase in the thermodynamic limit at $m=0$.

Interestingly, in the massless case, i.e., for $m=0$, a Haldane-like symmetry protected topological (SPT) phase~\cite{Senthil2015} appears between the VA and V0 phases. Similar to the Haldane phase in spin-1 chains~\cite{Haldane1983-2, Haldane1983-1, Affleck1987, DenNijs1989, Kennedy1992}, this SPT phase is characterized by the non-zero string order parameter, defined as
\begin{equation}
    O_S= \lim_{|k-l| \rightarrow \infty} |\langle \tilde{S}^z_{(k, j)}  e^{i \pi \sum_{n=k+1}^{l-1} \tilde{S}_{(n,j)}^z} \tilde{S}^z_{(l, j)}\rangle|,
\end{equation}
while the corresponding parity order parameter, defined as
\begin{equation}
    O_P= \lim_{|k-l| \rightarrow \infty} |\langle  e^{i \pi \sum_{n=k}^{l} \tilde{S}_{(n,j)}^z} \rangle|,
\end{equation}
vanishes. It should be emphasized that the string order parameter $O_S$ remains finite for all $J_y>0$ as a consequence of the local gauge constraint, whereas the parity order parameter vanishes exclusively in the SPT phase.

The SPT phase is protected by a $\mathbb{Z}_2 \times \mathbb{Z}_2$ symmetry of the Hamiltonian, with the corresponding operations:  
\begin{itemize}
    \item $\mathcal{C}$: particle–hole transformation, $\psi_{(i,j)} \leftrightarrow \psi_{(i,j)}^{\dagger}$ for all $(i,j)$, together with $S^{+} \leftrightarrow \sigma^y S^{+} \sigma^y$ on all links.  
    \item $\mathcal{R}$: staggered sign change, $\psi_{(i,j)} \leftrightarrow -\psi_{(i,j)}$ for sites with $(-1)^{i+j} = 1$, together with $S^{+} \leftrightarrow \sigma^z S^{+} \sigma^z$ on all links.
\end{itemize}

\begin{figure*}[tbh]
	\includegraphics[width=0.3\linewidth]{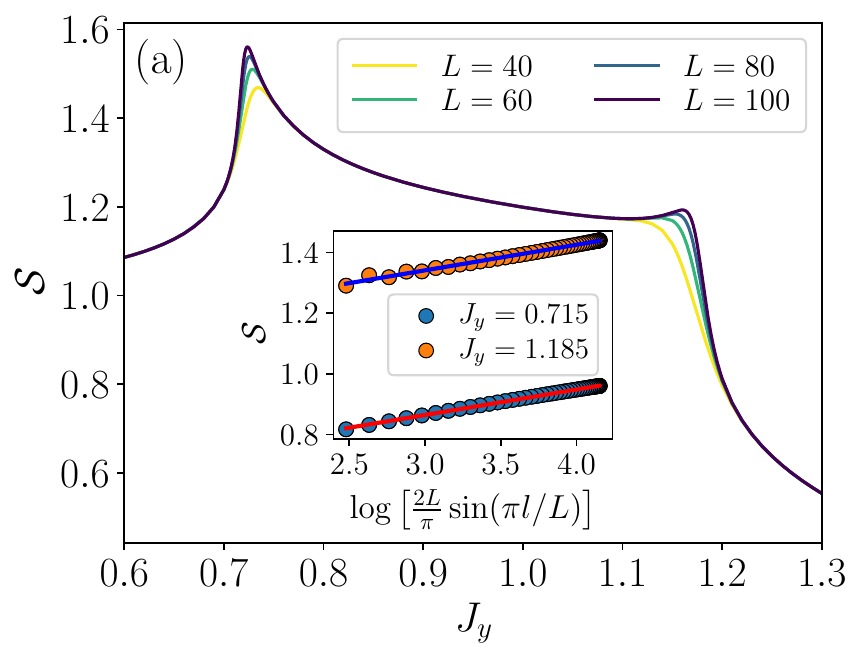}
    \includegraphics[width=0.3\linewidth]{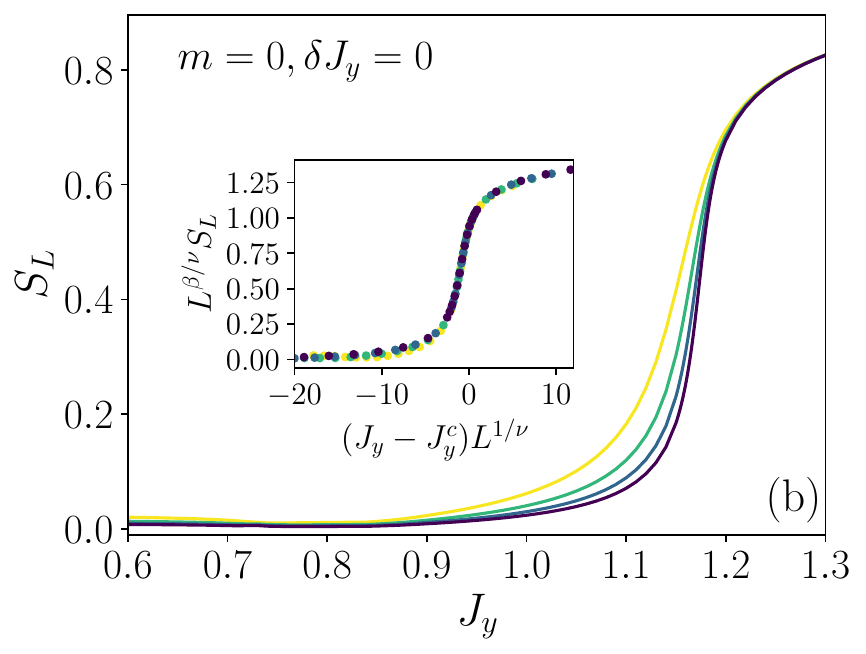}
    \includegraphics[width=0.3\linewidth]{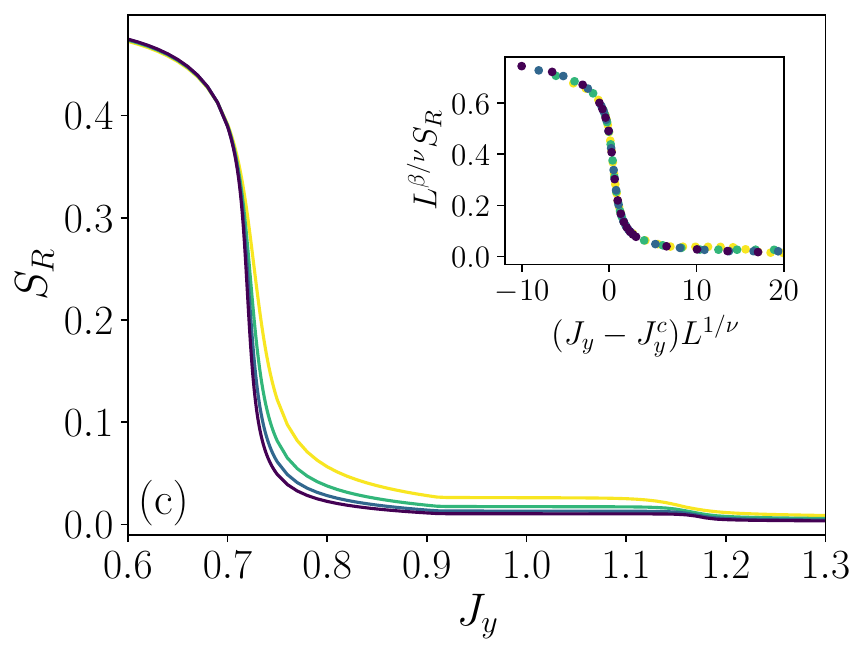}
	\caption{(Color online.) Entanglement entropy (a) at the middle of the ladder, leg magnetization $S_L$ (b), and rung magnetization $S_R$ (c) for the disorder-free system of lengths $L=40,60,80,100$ at zero fermion mass ($m=0$). The entropy profile clearly reveals two phase transitions separating the three gapped phases present at zero mass. The inset of (a) shows the critical entanglement scaling according to Eq.~\eqref{eq:cardy_calabrese} at the two critical points. Panels (b) and (c) demonstrate that $S_L$ and $S_R$ serve as order parameters for the SPT-to-V0 and VA-to-SPT transitions, respectively. The insets in (b) and (c) display the data collapse obtained from the universal finite-size scaling (Eq.~\eqref{eq:fss}).
	\label{fig:m=0_dJy=0}}
\end{figure*}

In this work, to distinguish and characterize different phase transitions, we also consider the von Neumann entanglement entropy of the ground state. For a block of length $l$ along the longitudinal direction, it is defined as
\begin{equation}
\mathcal{S}(l) = - \mathrm{Tr} \left[ \rho_l \ln \rho_l \right],
\end{equation}
where $\rho_l$ is the reduced density matrix of the block. Owing to the presence of gauge fields, several partitioning schemes are possible for computing the entanglement entropy; the scheme adopted here is illustrated in Fig.~\ref{fig:schem}(a). At the critical point, the entanglement entropy shows the universal scaling~\cite{callan_geometric_1994, vidal_PRL_2003, calabrese_entanglement_2004}:
\begin{equation}
    \mathcal{S}(l,L) = \frac{c}{6} \log \left[   \frac{2 L}{\pi} \sin(\pi l /L)\right] + b',
    \label{eq:cardy_calabrese}
\end{equation}
where $c$ in the central charge of the underlying conformal field theory (CFT) and $b'$ is a non-universal constant.

\begin{figure}[th]
\includegraphics[width=0.7\linewidth]{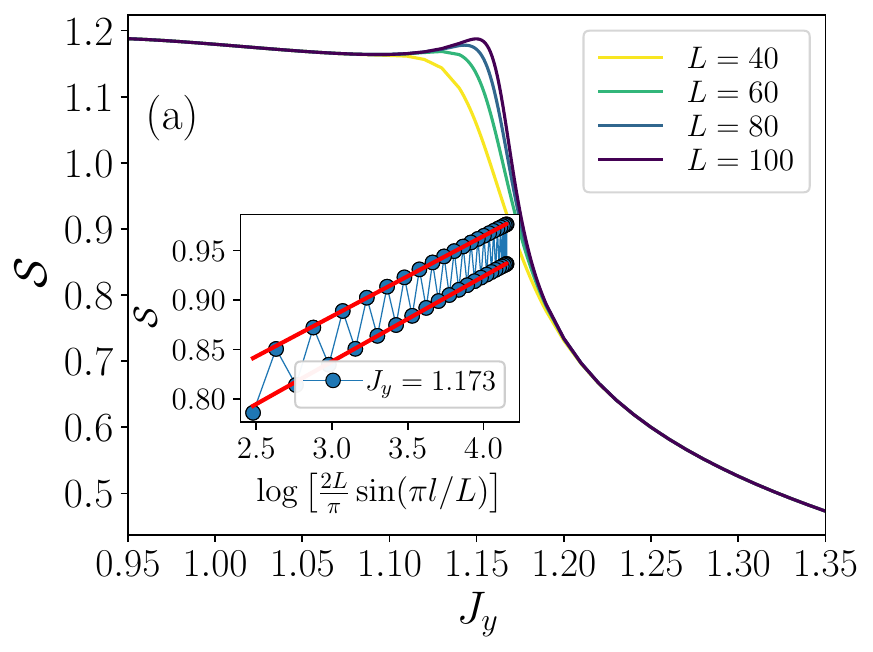}
\includegraphics[width=0.7\linewidth]{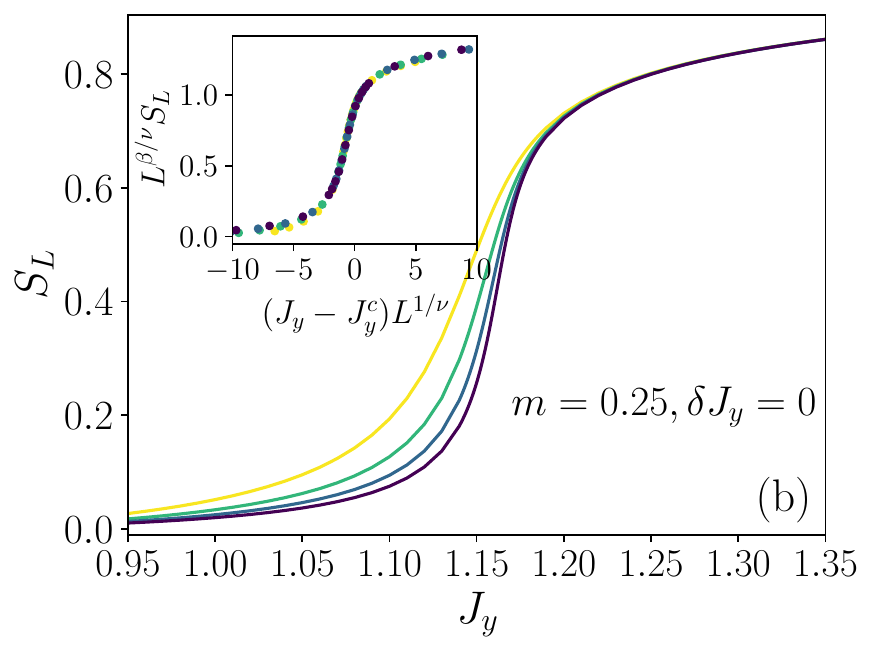}
	\caption{(Color online.) Entanglement entropy (a) at the center of the ladder and leg magnetization $S_L$ (b) in the disorder-free ladder of lengths $L=40,60,80,100$ at fermion mass $m=0.25$. The entropy indicates the presence of only two quantum phases, VA and V0, separated by a single transition. The inset of (a) shows the scaling of the block entanglement entropy with the chord length $\log\!\left[\tfrac{2L}{\pi} \sin(\pi l/L)\right]$. In contrast to the case in Fig.~\ref{fig:m=0_dJy=0}, the block-size dependence at criticality exhibits Friedel oscillations due to OBC. The central charge is extracted by fitting the scaling form separately to the two sublattices. The inset of (b) presents the finite-size scaling collapse of $S_L$ near the direct VA-to-V0 transition.}
	\label{fig:m=025_dJy=0}
\end{figure}

In this work, our primary goal is to investigate the stability and characterization of various phases and phase transitions in the presence of small but non-zero disorder.
For this purpose, we perform density-matrix renormalization group (DMRG)~\cite{White1992, White1993} simulations for finite ladders using matrix product state (MPS)~\cite{Schollwoeck2011, Orus2014} ansatz. 
Before addressing the disordered ladder, we first examine the disorder-free limit to establish a reference point.

\section{The phase diagram of the system without disorder}
\label{sec:phase-diagram}

The phase diagram of the system without disorder ($\delta J_y = 0$) is shown in Fig.~\ref{fig:phase-no-disorder}. Owing to the symmetry under a sign flip of the mass ($m \leftrightarrow -m$), it is sufficient to restrict the analysis to $m \ge 0$. The diagram is obtained by performing finite-size scaling for the entanglement entropy at the center of a system. For sufficiently large system sizes, the entanglement entropy exhibits clear kinks near the phase transitions. In the thermodynamic limit, these kinks coincide precisely with the phase transition points.

The entanglement entropy at the center of the ladder, along with the two magnetizations for various system sizes at zero mass ($m=0$), is shown in Fig.~\ref{fig:m=0_dJy=0}. As discussed earlier, the system hosts three gapped quantum phases: the VA phase at small $J_y$, the V0 phase at large $J_y$, and the SPT phase at intermediate $J_y$ (around $J_y \approx 1$). The entanglement entropy, plotted in Fig.~\ref{fig:m=0_dJy=0}(a), exhibits pronounced kinks at (or very close to) the two transition points. Exactly at these phase transitions, the entanglement entropy follows the critical scaling according to Eq.~\eqref{eq:cardy_calabrese}, as shown in the inset. Importantly, the extracted central charge $c$ is fully consistent with $1/2$ at both transitions, thereby providing clear evidence that the transitions belong to the Ising universality class.

The locations of the phase transitions are determined using the universal finite-size scaling hypothesis,
\begin{equation}
    Q(L) = L^{-\beta/\nu} \, f\!\left( (J_y - J_y^c) L^{1/\nu} \right),
    \label{eq:fss}
\end{equation}
where $J_y^c$ denotes the critical point, and $\nu$ and $\beta$ are the universal critical exponents. This analysis is applied to both order parameters, $S_L$ and $S_R$.
\tc{The details of the procedure used for the finite-size scaling analysis are presented in Appendix~\ref{app:fss}.}
The insets of Fig.~\ref{fig:m=0_dJy=0}(b) and (c) show the corresponding data collapses, demonstrating excellent agreement with the scaling form. The values of the extracted critical parameters are summarized in Table~\ref{tab:disorderless}.
\tc{The estimated values of the correlation length exponent, $\nu \approx 1$, are in excellent agreement with the Ising universality class for both the transitions.}

\begin{table}[htb]
	\centering
	\begin{tabular}{|c|c|c|c|}
		\hline
		\ & \ VA-to-SPT \ & \ SPT-to-V0 \ & \ VA-to-V0 \\
		\ & \ ($m=0$) \ & \ ($m=0$) \ & \ ($m=0.25$) \\
		\hline
		$J_y^c$ & 0.715(1) & 1.186(1) & 1.172(1) \\
		$\nu$ & 1.03(2) & 1.08(5) & 1.09(6) \\
        $\tc{\beta}$ & \tc{0.101(7)} & \tc{0.106(7)} & \tc{0.108(9)} \\
        $c$ & 0.50(1) & 0.50(1) & 0.50(2) \\
		\hline
	\end{tabular}

	\caption{Critical parameters extracted for the three types of phase transitions: VA-to-SPT and SPT-to-V0 at $m=0$, and the direct VA-to-V0 transition at $m=0.25$. 
    The critical coupling $J_y^c$ and \tc{the exponents $\nu$, $\beta$} are obtained from finite-size scaling of the corresponding order parameter (leg or rung magnetization), while the central charge $c$ is obtained from the scaling of entanglement entropy. The central charge $c$ and correlation-length exponent $\nu$ are consistent with the Ising universality class in all cases.}	
	\label{tab:disorderless}	
\end{table}
\tc{We emphasize that the estimated values of the order-parameter exponent $\beta$ do not coincide with the Ising value $\beta = 1/8$. However, such deviations do not preclude Ising criticality, since $\beta$ is operator dependent. For example, in the Ising model, the $n$th power of the magnetization scales with an effective exponent $\beta = n/8$. In the present case, due to the absence of an exact mapping between the quantum link ladder model and the Ising theory, we cannot unambiguously relate the rung or longitudinal magnetization to the primary magnetization operator of the Ising conformal field theory. Consequently, the extracted values of $\beta$, reported in Table~\ref{tab:disorderless}, are not expected to match $1/8$.
That is why, the universality class is more reliably identified through operator-independent quantities, such as the correlation length exponent $\nu$ and the central charge $c$, which together are sufficient to conclude here that the transitions belong to the Ising universality class.}

Figure~\ref{fig:m=025_dJy=0} investigates the direct VA-to-V0 phase transition at finite mass $m=0.25$, where we present the mid-ladder entanglement entropy $\mathcal{S}$ together with the leg magnetization $S_L$. In contrast to the massless case, the spatial dependence of the entanglement entropy at criticality displays pronounced Friedel oscillations \cite{Friedel58, Laflorencie06, Eggert95, Jiang18}, a consequence of the OBC (inset of Fig.~\ref{fig:m=025_dJy=0}(a)). The universal scaling of $\mathcal{S}$ with the chord length $\log\!\left[\tfrac{2L}{\pi} \sin(\pi l/L)\right]$ is therefore fitted separately on the two sublattices, yielding a central charge $c=1/2$ (Table~\ref{tab:disorderless}), in agreement with the Ising universality class. The location of the transition $J_y^c$ is independently determined from the finite-size scaling collapse of the leg magnetization $S_L$ (inset of Fig.~\ref{fig:m=025_dJy=0}(b)), which also provides the correlation-length exponent $\nu=1$. Taken together, these results establish that the direct VA-to-V0 transition at $m=0.25$ is a continuous transition governed by the Ising fixed point.  

In the following sections, we will discuss how the phase diagram of the system changes in the presence of disorder. As we will see, due to the quasi-one-dimensional geometry of the ladder, it is important which part of the Hamiltonian is affected by the disorder.

\section{The bond disorder on the rungs}
\label{sec:disorder}

Let us first comment, however, that adding a diagonal on-site disorder (as often studied in cold atom community) leads to somewhat obvious cancellation of the SPT phase as the mass can become non-zero at different sites. It is more rewarding to address the bond disorder.  For that reason, we consider first the behavior of the quantum link ladder system in the presence of quenched disorder in the rung couplings $J_{yi}$. We aim to investigate how the disorder affects the quantum phases identified in the clean limit, and to understand the nature of the associated phase transitions.

\subsection{Non-zero fermion mass ($m>0$)}\label{subsec:nonzero-mass}

We start our discussion with the technically simpler case of a non-zero fermion mass. As in the previous section, we focus on the example $m=0.25$, which allows for a direct comparison of the ground-state properties in the absence and in the presence of disorder. Throughout this part, we mainly concentrate on weak disorder with strength $\delta J_y = 0.1$, and later we comment on the fate of the system at stronger disorder values. To obtain statistically reliable results for disorder-averaged observables, we consider several disorder realizations $D$ chosen such that $L D = 40000$, which ensures comparable statistical accuracy across different system sizes.

\begin{figure}[tbh]
    \includegraphics[width=0.7\linewidth]{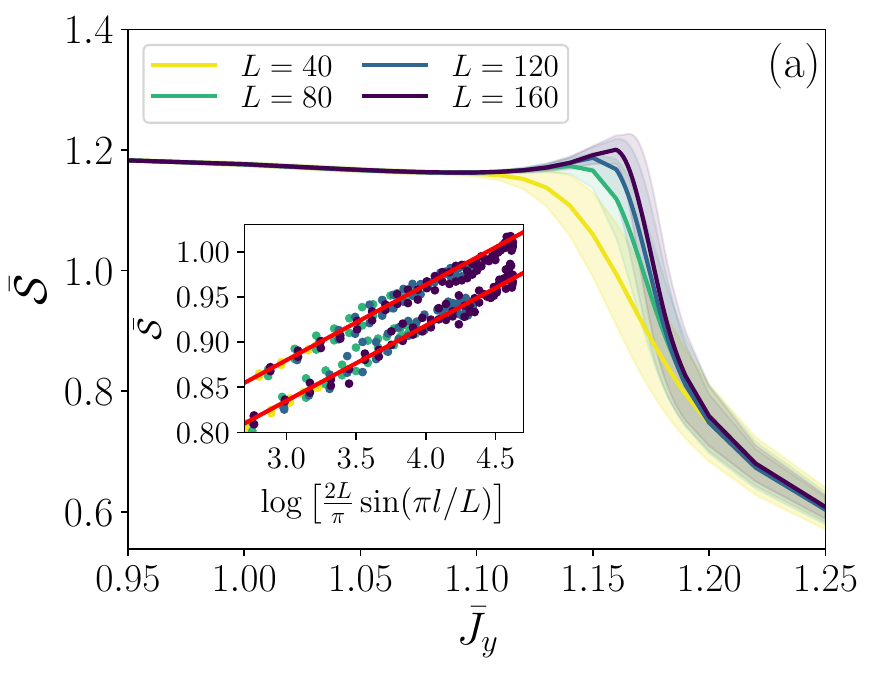}
    \includegraphics[width=0.7\linewidth]{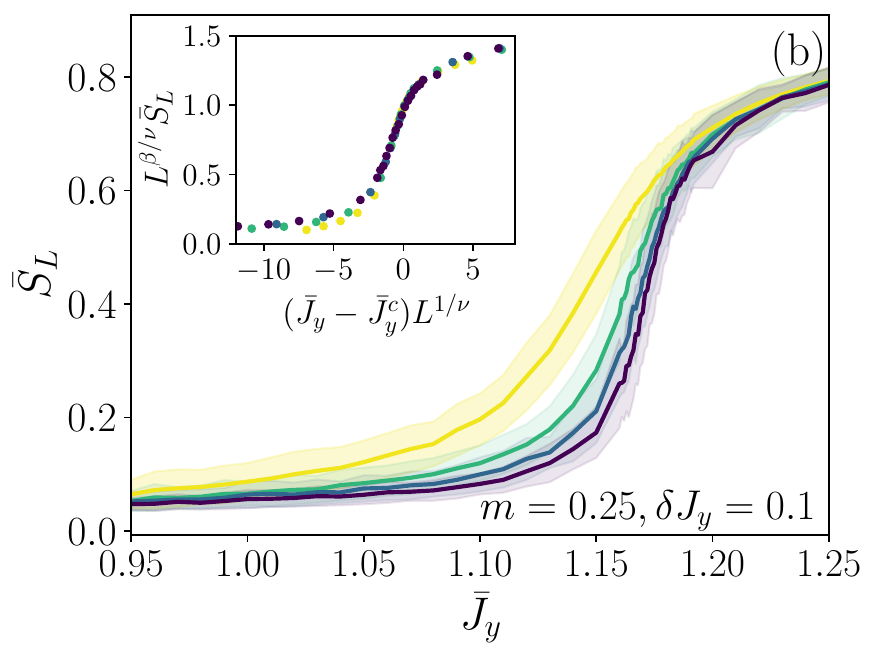}
	\caption{(Color online.) Disorder-averaged entanglement entropy $\bar{\mathcal{S}}$ at the center of the ladder and disorder-averaged leg magnetization $\bar{S}_L$ for system sizes $L=40, 80, 120, 160$ at fermion mass $m=0.25$ and disorder strength $\delta J_y=0.1$. 
    Shaded regions indicate error bars obtained from sample averaging. 
    The entropy clearly signals a direct VA-to-V0 transition in the presence of disorder. 
    The inset of (a) shows the logarithmic scaling of $\bar{\mathcal{S}}$ with chord length at the critical point. The Friedel oscillations due to OBC are clearly observed.
    The critical coupling $\bar{J}_y^c$ and the associated universal exponents are determined from the finite-size scaling collapse of $\bar{S}_L$, as displayed in the inset of (b).}
	\label{fig:m=025_dJy=01}
\end{figure}

In the presence of disorder in $J_{yi}$ with strength $\delta J_y = 0.1$, the structure of the phase diagram at non-zero mass remains essentially unchanged. The system still supports two gapped phases: the VA phase and the V0 phase.  
The disorder-averaged entanglement entropy $\bar{\mathcal{S}}$ and the disorder-averaged leg magnetization $\bar{S}_L$ display clear signatures of a direct transition between these two phases.  
In the VA phase, $\bar{S}_L$ is small, while in the V0 phase it remains finite, with the transition point shifted slightly compared to the clean case of $\delta J_y = 0$. Furthermore, the numerical data indicate that in the thermodynamic limit, the VA phase develops a small but finite residual magnetization, $\lim_{L \rightarrow \infty} \bar{S}_L > 0$, attributable to disorder-induced defects.  
The behaviors of $\bar{\mathcal{S}}$ and $\bar{S}_L$ are shown in Fig.~\ref{fig:m=025_dJy=01} as functions of the average rung tunneling amplitude $\bar{J}_y$ across the phase transition. Finite-size scaling analyses at the transition further confirm the robustness of the Ising universality class: the scaling of $\bar{\mathcal{S}}$ according to Eq.~\eqref{eq:cardy_calabrese} yields a central charge $c=1/2$ (inset of Fig.~\ref{fig:m=025_dJy=01}(a)), while the finite-size data collapse of $\bar{S}_L$ gives the critical exponent $\nu=1$ (inset of Fig.~\ref{fig:m=025_dJy=01}(b)). The corresponding exact extracted values are summarized in Table~\ref{tab:disorder}, establishing that the VA-to-V0 transition remains consistent with the Ising universality class even in the presence of weak disorder. Interestingly, we observe that the critical $J_y^c=1.177$ value obtained corresponds to the position in $\bar J_y$  of the inflection point of the disorder-averaged leg magnetization $\bar S_L$ and the rung magnetization $\bar{S}_R$.

\begin{figure}
    \centering
    \includegraphics[width=0.49\linewidth]{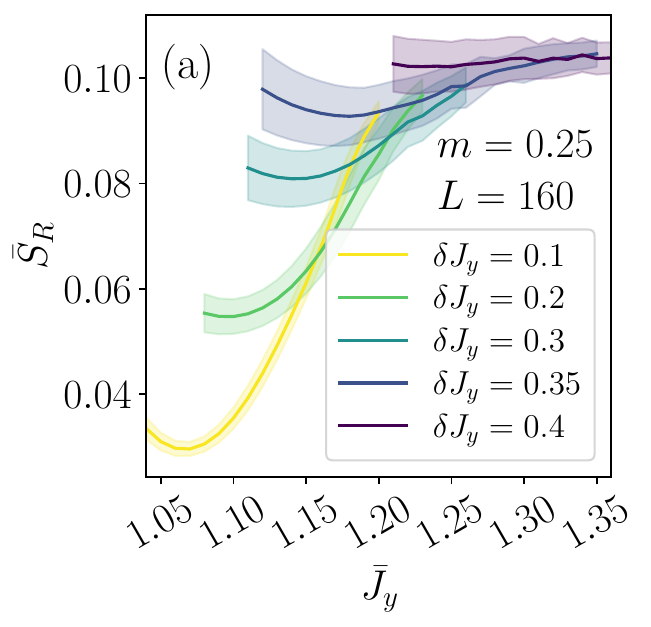}
    \includegraphics[width=0.49\linewidth]{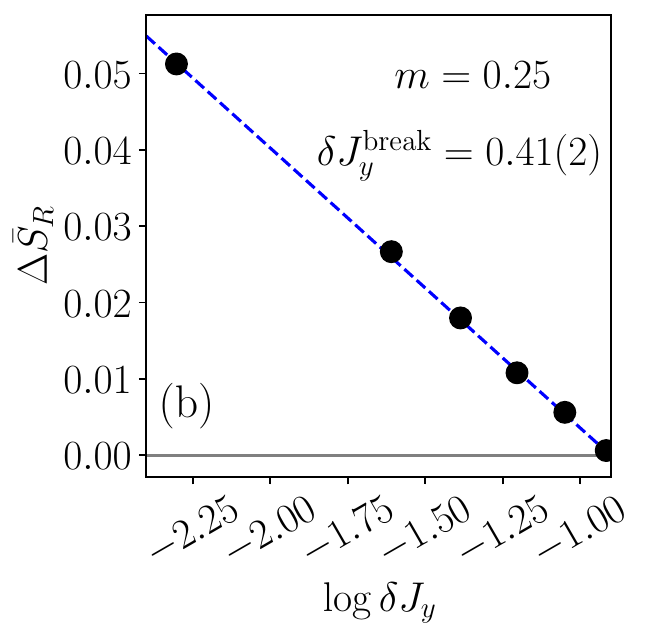}
    \caption{(Color online.) (a) Disorder-averaged rung magnetization $\bar S_R$ for system size $L=160$ at fermion mass $m=0.25$ for different disorder strengths $\delta J_y$ across the direct VA-to-V0 transition or crossover.
    (b) The depth of the minima in $\bar S_R$ appearing near the transition for different disorder amplitudes in a log-linear scale. The depth is measured by the difference $\Delta \bar S_R$ between the $\bar S_R$ values measured at the inflection point and at the minimum. The $\delta J_y^{\text{break}} = 0.41(2)$ denotes the disorder strength at which the $\bar J_y$ curve flattens out and the minimum disappears.
    }
    \label{fig:deltaSR}
\end{figure}

The structure of the phase diagram, along with the Ising transition at finite mass, remains robust up to disorder strengths of about $\delta J_y \simeq 0.4$. This can be observed, e.g., by comparing the shapes of the disordered-averaged rung magnetizations $\bar S_R$ for different disorder amplitudes as depicted in Fig.~\ref{fig:deltaSR}(a). At a smaller disorder strength, $\bar S_R$ shows a minimum just before the transition, followed by an inflection point at the transition.
The pronounced minimum appearing at smaller disorder amplitudes flattens out and eventually disappears at $\delta J_y \simeq 0.4$ as disorder strength is slowly increased.
Beyond $\delta J_y \simeq 0.4$, the behavior changes qualitatively. Both the disorder-averaged entanglement entropy and the disorder-averaged magnetizations exhibit only weak dependence on the system size $L$ and no clear signatures of a phase transition.
This suggests that beyond this threshold disorder strength, the distinction between the VA and V0 phases, as well as the Ising transition between them, is lost, and the system no longer supports well-defined long-range ordered phases. 
Instead, the ground state is characterized by strong spatial inhomogeneity, where local regions fluctuate between vortex–antivortex–dominated and vortex–zero–flux–dominated configurations, preventing the emergence of a sharp phase transition.

To quantitatively locate the position of this threshold disorder strength, we estimate the depth of the minima in $\bar S_{R}$ for different disorder
amplitudes. To this end, we measure the difference, $\Delta \bar S_R$, between the $\bar S_R$ value at the inflection point of the curve (which indicates the position of the crossover/transition) and the $\bar S_R$ value at the minimum. The results, plotted in log-linear scale in Fig.~\ref{fig:deltaSR}(b), allow us to estimate the position when the $\bar S_R$ curve flattens out at $\delta J_y^{\rm break} \approx 0.41$. For a stronger disorder, the phase transition between phases disappears.

\begin{table}[htb]
	\centering
	\begin{tabular}{|c|c|c|c|}
		\hline
		\ & \ VA-to-V0 \ & VA-to-SPT \ & \ SPT-to-V0 \\
		\ & \ ($m=0.25$) \ & \ ($m=0$) \ & \ ($m=0.0$) \\
		\hline
		$J_y^c$ & 1.178(5) & 0.717(8) & 1.186(5) \\
		$\nu$ & 1.08(12) & 1.06(14) & 1.09(10) \\
        $\tc{\beta}$ & \tc{0.128(12)} & \tc{0.123(14)} & \tc{0.129(11)} \\
        $c$ & 0.51(2) & 0.50(1) & 0.51(1) \\
		\hline
	\end{tabular}	
	\caption{Critical parameters extracted for the three types of phase transitions in the presence of disorder with strength $\delta J_y=0.1$: the direct VA-to-V0 transition at $m=0.25$, and VA-to-SPT and SPT-to-V0 at $m=0$. 
    The critical coupling $\bar{J}_y^c$ and \tc{the exponents $\nu$, $\beta$} are obtained from finite-size scaling of the corresponding order parameter (disorder-averaged leg or rung magnetizations), while the central charge $c$ is obtained from the scaling of entanglement entropy.}	
	\label{tab:disorder}	
\end{table}

\subsection{Field theoretic argument for the stability of the Ising transition} \label{subsubsec:FT}

It is instructive to examine the seemingly \textit{unusual} stability of the Ising criticality in the presence of disorder from the perspective of the Harris criterion. The Harris criterion~\cite{Harris74} states that disorder is a relevant perturbation at a clean critical point if $d\nu < 2$, in which case even an arbitrarily weak disorder destabilizes the transition. In the present case, $d=1$ and $\nu=1$, and one would therefore naively expect that, in the presence of disorder, the Ising transition of the clean system should flow to a \textit{strong-disorder fixed point}~\cite{IGLOI2005}. The observed stability of the Ising criticality thus appears to violate the Harris criterion. Below, we provide a phenomenological field-theoretic rationale for this apparent contradiction for non-zero mass $m$.

When all $J_{y i}=0$, the QLM on a two-leg ladder decouples into two identical QLM chains. Each chain independently undergoes an Ising transition~\cite{Rico2014}, which is described at low energies by the standard real (Majorana) fermion field theory with the Hamiltonian~\cite{Gogolin2004}
\begin{align}
    H = \sum_{j=0,1} i \int dx \left[ v\, \eta_j(x)\,\partial_x \zeta_j(x) + \mu\, \eta_j(x)\,\zeta_j(x) \right],
\end{align}
where $\eta_j$ and $\zeta_j$ denote the pair of Majorana fields on the $j$th leg, $v$ is the sound velocity, and $\mu$ is the Majorana mass. At the Ising critical point, the mass $\mu$ vanishes, yielding two copies of the Ising CFT with total central charge $c = 1/2 + 1/2 = 1$.

In the presence of a uniform interleg coupling, $J_{y i} = J_y$, the Majorana fields on the two legs become coupled. The low-energy theory can then be expressed in terms of symmetric and antisymmetric combinations of the original fields (see, e.g.,~\cite{Tsitsishvili2022}):
\begin{align}
    H &= i \int dx \left[ v_S\, \eta_S(x)\,\partial_x \zeta_S(x) + \mu_S\, \eta_S(x)\,\zeta_S(x)\right] \nonumber \\
      & + i \int dx \left[ v_A\, \eta_A(x)\,\partial_x \zeta_A(x) + \mu_A\, \eta_A(x)\,\zeta_A(x)\right],
    \label{eq:maj_field}
\end{align}
where
\begin{align}
\eta_S = \frac{\eta_0 + \eta_1}{\sqrt{2}}, \quad
\eta_A = \frac{\eta_0 - \eta_1}{\sqrt{2}}, \nonumber \\
\zeta_S = \frac{\zeta_0 + \zeta_1}{\sqrt{2}}, \quad
\zeta_A = \frac{\zeta_0 - \zeta_1}{\sqrt{2}},    
\end{align}
and $v_{S,A}$ and $\mu_{S,A}$ are the corresponding velocities and Majorana masses. We should note that once an interchain coupling is introduced, e.g., for $J_y \neq 0$, the symmetric and antisymmetric sectors generically acquire different velocities and masses. 
Since the clean QLM ladder exhibits only a single Ising critical point with central charge $c=1/2$ at non-zero (complex) fermion mass $m$, it follows that only one sector, specifically the antisymmetric sector due to the symmetry under leg-exchange in conjunction with $m \rightarrow -m$, goes through the Ising transition as $\mu_A$ changes sign, while the symmetric sector remains gapped with $\mu_S \neq 0$.

The stability of the Ising transition in the presence of disorder can now be understood if the disorder couples exclusively to the symmetric sector in the effective field theory~\eqref{eq:maj_field}. Since the symmetric sector remains gapped throughout, disorder does not affect the low-energy critical theory governed solely by the massless antisymmetric Majorana fields. Consequently, the Ising criticality remains stable even in this disordered one-dimensional system, effectively bypassing the Harris criterion.

Although a fully controlled microscopic derivation of the effective field theory~\eqref{eq:maj_field} for the ladder system is currently lacking, this phenomenological description is fully consistent with our numerical results and provides a natural explanation for the observed robustness of the Ising criticality in the disordered one-dimensional QLM ladder.

\subsection{Zero fermion mass ($m=0$)}\label{subsec:zero-mass}

\begin{figure}
    \centering
    \includegraphics[width=0.9\linewidth]{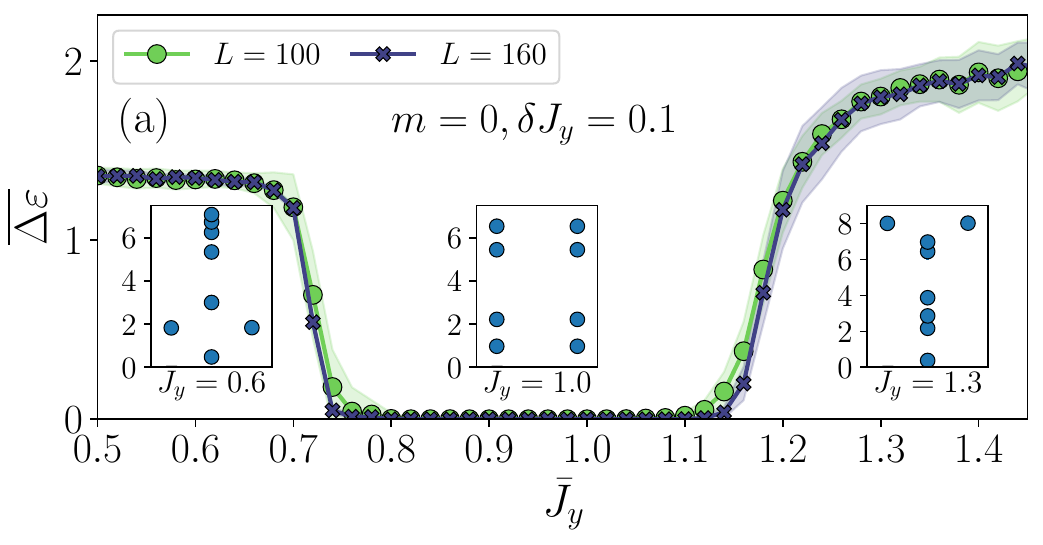}
    \includegraphics[width=0.9\linewidth]{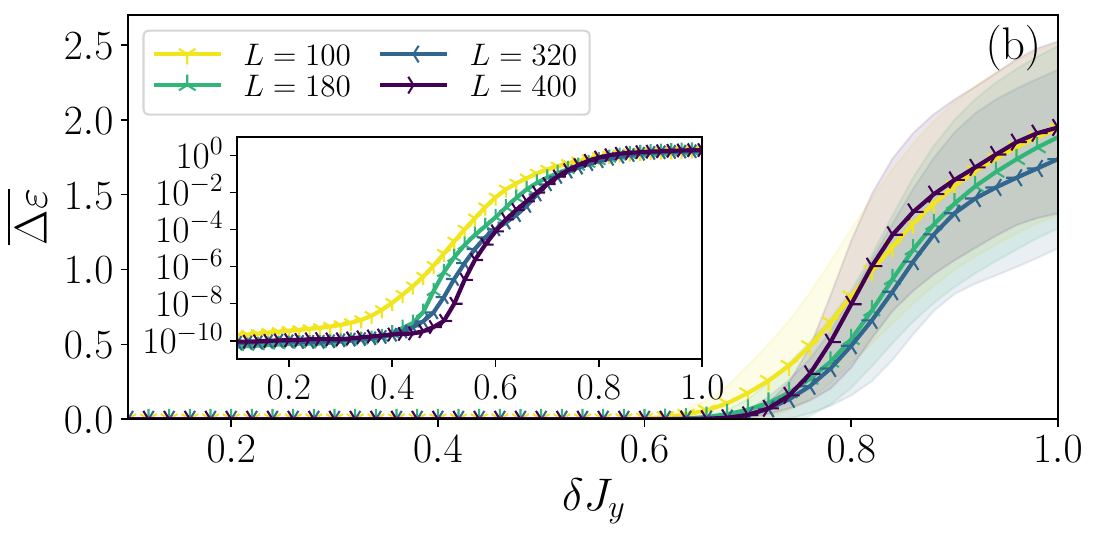}
    \caption{(Color online.) (a) Disorder-averaged entanglement gap $\overline{\Delta \varepsilon}$ at the center of the ladder for system sizes $L=100, 160$ at zero mass and disorder strength $\delta J_y=0.1$. Shaded regions denote error bars. (Insets of (a)) Entanglement spectrum in the three phases: VA ($\bar{J}_y = 0.6$), SPT ($\bar{J}_y = 1$), and V0 ($\bar{J}_y = 1.3$) for $L=160$. In the insets of panel (a), error bars are smaller than or comparable to the symbol sizes.
    (b) Disorder-averaged entanglement gap $\overline{\Delta \varepsilon}$ at the center of the ladder with increasing disorder strength $\delta J_y$ for different system-sizes at $\bar{J}_y = 1$ and $m=0$. (Inset of (b)) Same as in the main panel but the vertical axis is presented in log-scale.
    }
    \label{fig:egap_dJy=01}
\end{figure}

In the previous parts, we have demonstrated that for weak disorder the two phases present in the clean limit, together with the intervening Ising transition, remain robust and are only destroyed at sufficiently large disorder strength. In this section, we turn to the massless limit ($m=0$), where the central question concerns the stability of the SPT phase that exists in the clean case. 

\begin{figure*}[tbh]
	\includegraphics[width=0.24\linewidth]{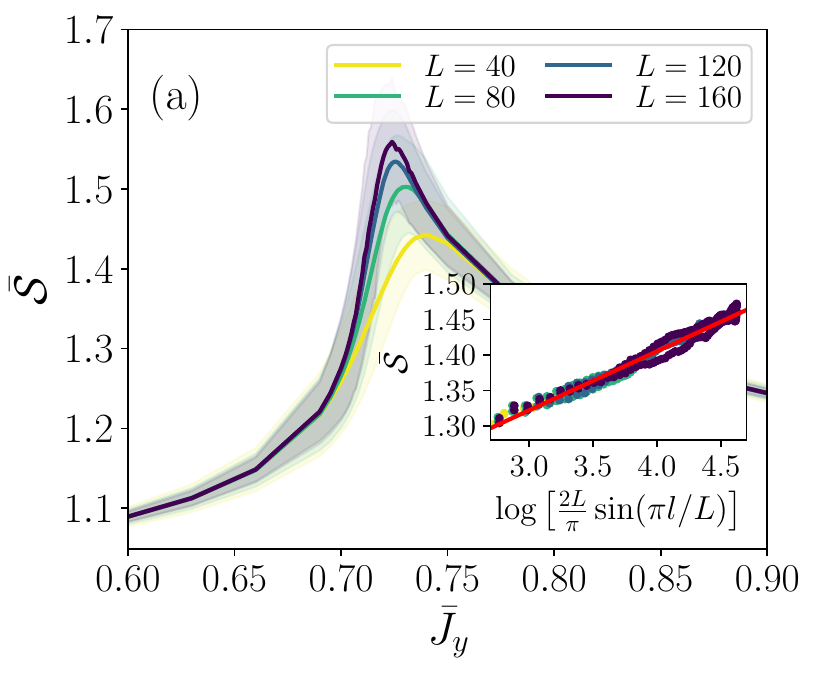}
    \includegraphics[width=0.24\linewidth]{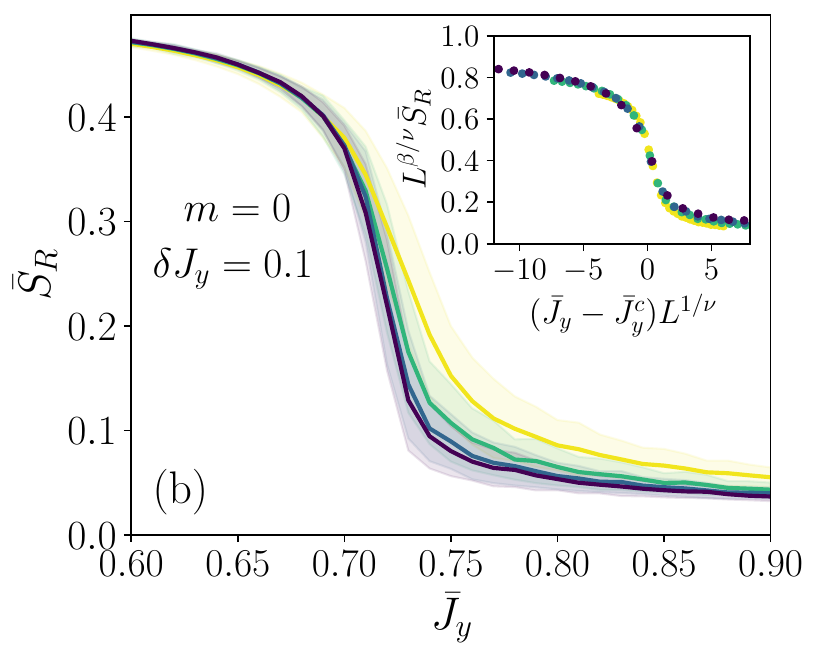}
    \includegraphics[width=0.24\linewidth]{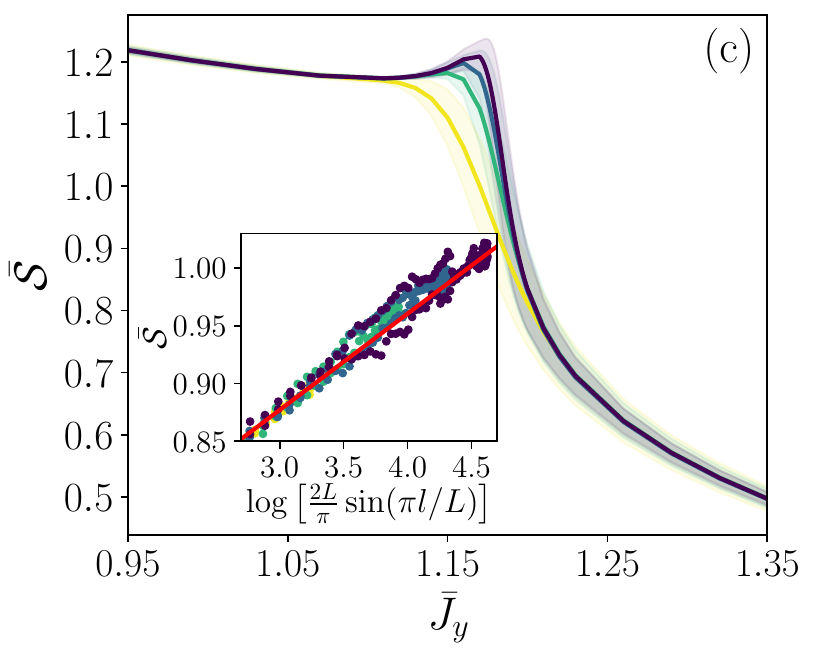} 
    \includegraphics[width=0.24\linewidth]{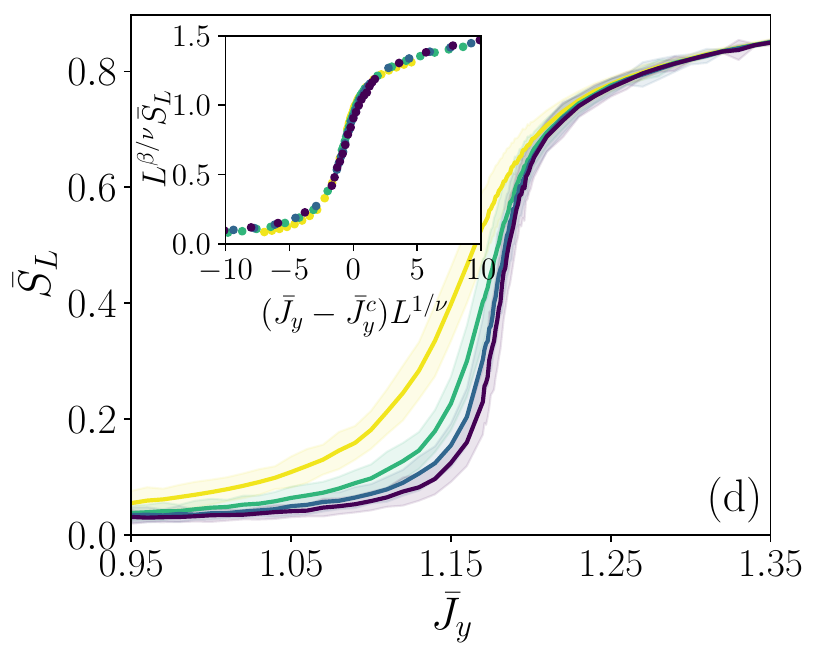}
	\caption{(Color online.) Disorder-averaged entanglement entropy $\bar{\mathcal{S}}$ in the middle of the ladder (a),(c), the rung magnetization $\bar{S}_R$ (b), and leg magnetization $\bar{S}_L$ (d) for system sizes $L=40, 80, 120, 160$ at zero fermion mass and disorder strength $\delta J_y=0.1$. 
    Panels (a) and (b) correspond to the VA-to-SPT transition, while (c) and (d) correspond to the SPT-to-V0 transition. 
    Finite-size scaling of the disorder-averaged magnetizations (insets of panels (b) and (d)) yields the critical exponent $\nu = 1$ in both cases, confirming the stability of the Ising transition in presence of disorder. 
    Furthermore, the logarithmic scaling of the disorder-averaged entropy at the transitions (insets of panels (a) and (c)), is consistent with Ising universality, yielding a central charge $c = 1/2$. 
    All other details are the same as in Fig.~\ref{fig:m=025_dJy=01}.
    }
	\label{fig:m=0_dJy=01}
\end{figure*}

It is important to note that, even in the presence of disorder in the rung tunneling amplitudes $J_{yi}$, the underlying $\mathbb{Z}_2 \times \mathbb{Z}_2$ symmetry associated with the $\mathcal{C}$ and $\mathcal{R}$ transformations remains intact, and therefore the symmetry protection of the SPT phase is, in principle, preserved. To examine the stability of the SPT phase, we analyze the entanglement spectrum $\{\varepsilon_i\}$ of a block of length $l$, defined as
\begin{equation}
    \varepsilon_i = -\log \lambda_i,
\end{equation}
where $\{\lambda_i\}$ are the eigenvalues of the reduced density matrix $\rho_l$ of the block, ordered in decreasing magnitude. The entanglement gap, 
\begin{equation}
    \Delta \varepsilon = \varepsilon_1 - \varepsilon_0,
\end{equation}
is defined as the difference between the two lowest entanglement energy levels. For one-dimensional (or quasi-one-dimensional) systems, the topological nature of the SPT phase is reflected in a characteristic twofold degeneracy of the entanglement spectrum~\cite{Pollmann2010, Pollmann2012}, corresponding to a vanishing entanglement gap $\Delta \varepsilon = 0$.

In Fig.~\ref{fig:egap_dJy=01}(a), we present the disorder-averaged entanglement gap $\overline{\Delta \varepsilon}$ evaluated at the center of the system as a function of the average rung tunneling $\bar{J}_y$, for disorder strength $\delta J_y = 0.1$ and mass $m=0$, with system sizes $L=100$ and $160$. In the region near $\bar{J}_y = 1$, the entanglement gap vanishes, providing clear evidence that the SPT phase remains stable against weak disorder. The insets display the lowest eight entanglement energy levels in the three representative phases, illustrating the characteristic twofold degeneracy of the spectrum at $\bar{J}_y=1$, in contrast to the non-degenerate spectra in the adjacent phases.
Figure~\ref{fig:egap_dJy=01}(b) shows the disorder-averaged entanglement gap $\overline{\Delta \varepsilon}$ at $\bar{J}_y = 1$ and $m=0$ as a function of disorder strength for different system-sizes. Clearly, as evident from the figure, the SPT phase remains robust even with increasing disorder strength and is destroyed only beyond $\delta J_y \gtrsim 0.5$, where the entanglement gap opens up, signaling the loss of topological protection.

In Fig.~\ref{fig:m=0_dJy=01}, we present the disorder-averaged entanglement entropy $\bar{\mathcal{S}}$ along with the rung and leg magnetizations $\bar{S}_R$ and $\bar{S}_L$ for different system sizes at $m=0$ and $\delta J_y=0.1$. Panels (a),(b) correspond to the VA-to-SPT transition, while (c),(d) show the SPT-to-V0 transition. In both cases, $\bar{\mathcal{S}}$ exhibits pronounced peaks at the critical couplings, accompanied by sharp changes in the corresponding magnetizations that serve as order parameters. Finite-size scaling of $\bar{S}_R$ and $\bar{S}_L$ (insets of (b),(d)) yields $\nu \approx 1$, fully consistent with the Ising universality class.  
Moreover, the logarithmic scaling of $\bar{\mathcal{S}}$ according to Eq.~(\ref{eq:cardy_calabrese}), shown for the SPT-to-V0 transition (inset of (c)), yields a central charge $c=1/2$, again characteristic of Ising criticality. Together, these results confirm that the two transitions remain continuous and preserve their universality class even in the presence of weak disorder. The precisely extracted critical parameters are summarized in Table~\ref{tab:disorder}.

We note that the stability of this pair of Ising critical points at zero fermion mass $m$ cannot be explained within the phenomenological field-theoretic description of Sec.~\ref{subsubsec:FT}, since neither these two Ising transitions nor the associated SPT phase exists in the decoupled limit to begin with. The mechanism underlying the stability of the SPT phase and the two Ising transitions in this regime is therefore qualitatively different from that in the massive case. This distinction becomes clear when disorder is introduced in the tunneling amplitudes $J_{xij}$ along the legs of the ladder, as we discuss in the following section.

\section{Bond disorder on the Legs}
\label{sec:disx}

\begin{figure}
    \centering
    \hspace{0.13cm}\includegraphics[width=0.49\linewidth]{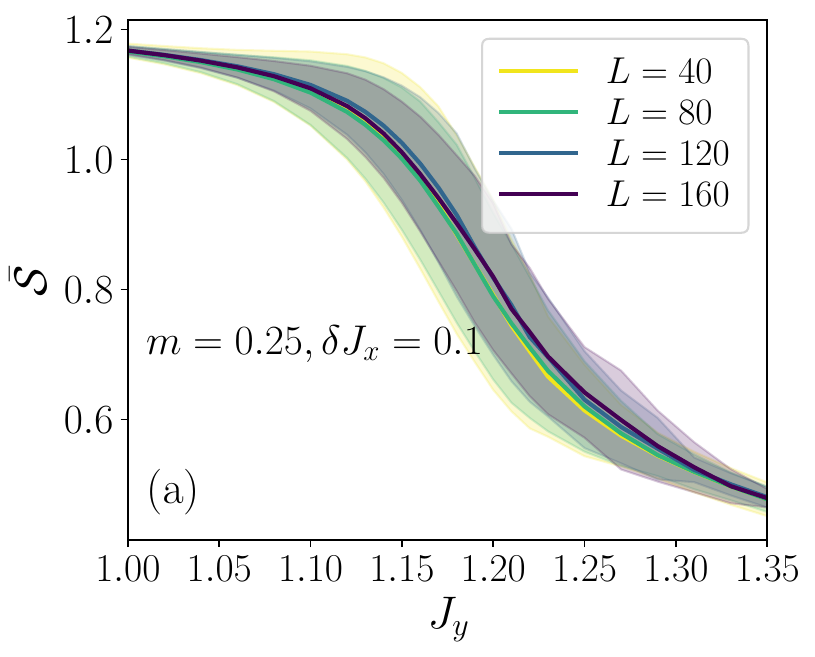}\hspace{-0.06cm}
    \includegraphics[width=0.49\linewidth]{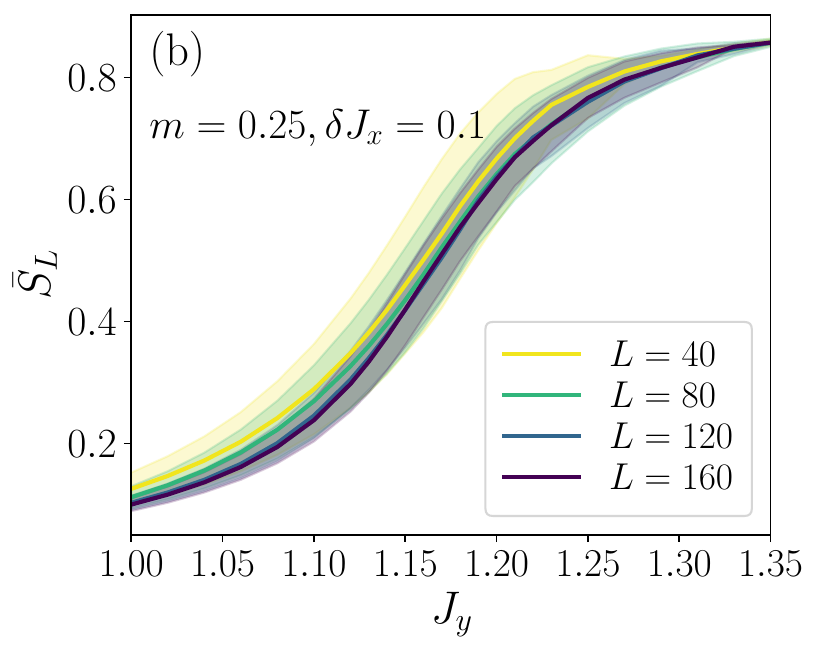}
    \includegraphics[width=0.49\linewidth]{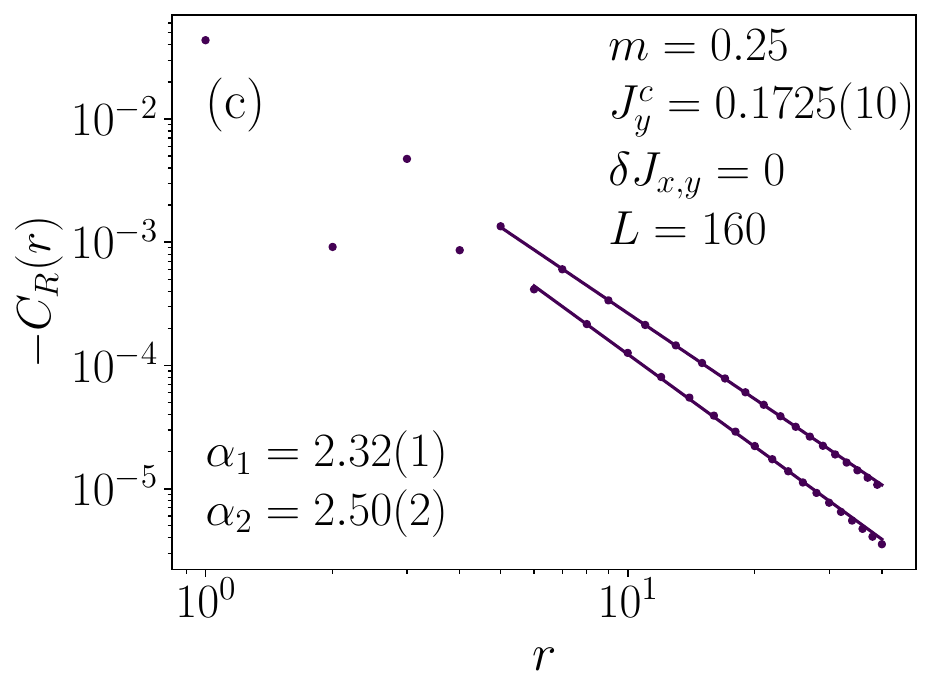}\hspace{-0.04cm}
    \includegraphics[width=0.49\linewidth]{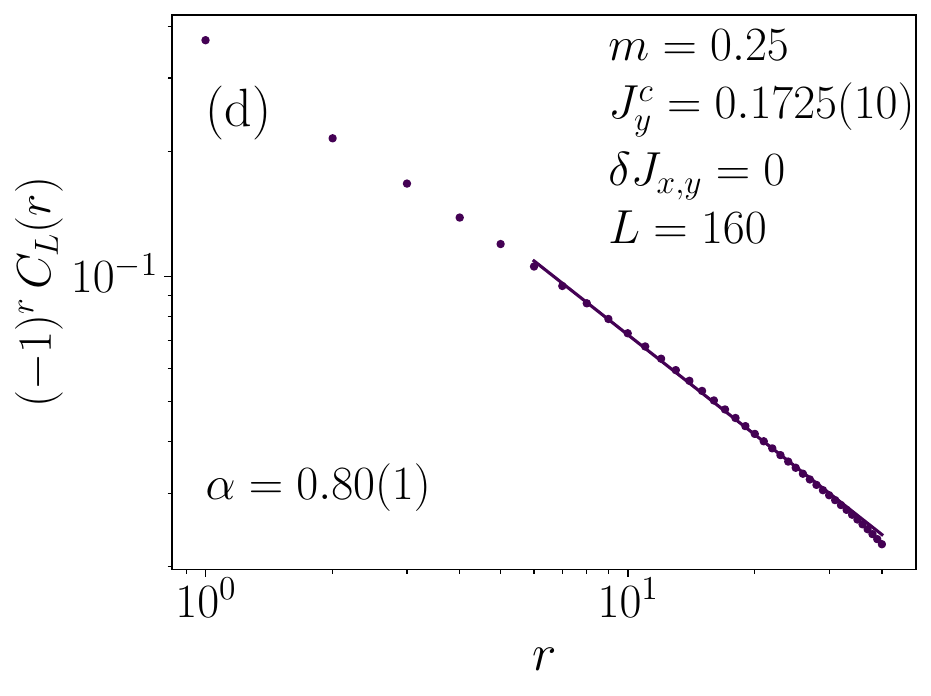}
    \caption{(Color online.) Top: Disorder-averaged entanglement entropy $\bar{\mathcal{S}}$ in the middle of the ladder (a) and leg magnetization $\bar{S}_L$ (b) in presence of bond disorder on the legs for system sizes $L=40, 80, 120, 160$ at non-zero fermion mass $m=0.25$ and disorder strength $\delta J_x=0.1$ as the rung tunneling amplitude $J_y$ is varied. The average $\bar{J}_x$ is fixed to 1. All other details are the same as in Fig.~\ref{fig:m=025_dJy=01}. \tc{Bottom: The correlation functions $C_{R,L}(r)$ (see Eq.~(\ref{eq:R_and_L_correlators})) of `local' magnetizations in the rung (c) and in the leg (d) directions for the non-disordered system of size $L=160$ with fermion mass $m=0.25$ at the phase transition. These correlations decay \jz{as a power law $r^{-\alpha}$ with $\alpha$ values indicated in the panels showing that} a rung/leg disorder is an irrelevant/relevant perturbation.}}
    \label{fig:m=0.25_dJx=0.1}
\end{figure}

The field-theoretic argument of Sec.~\ref{subsubsec:FT} predicts that, in the presence of disorder along the legs, the Ising transition in the massive regime ($m \neq 0$) of the clean system does not persist, since in the decoupled limit disorder is already a relevant perturbation. \tc{For demonstration, we compare the power-law decay of correlations of `local' magnetizations in the rung/leg direction:}
\tc{
\begin{align}
    C_R(r) =& \langle S_{(i,0),(i,1)}^z S_{(i+r,0),(i+r,1)}^z \rangle \nonumber \\ &- \langle S_{(i,0),(i,1)}^z \rangle \langle S_{(i+r,0),(i+r,1)}^z \rangle, \label{eq:R_and_L_correlators} \\
    C_L(r) =&\langle \tilde{S}_{(i,j)}^z \tilde{S}_{(i+r,j)}^z \rangle - \langle \tilde{S}_{(i,j)}^z \rangle \langle \tilde{S}_{(i+r,j)}^z \rangle,\nonumber
\end{align}
}
\tc{with distance at the clean phase transition point ($J_y \simeq 1.1725$, $m=0.25$) for a large system size $L=160$: $C_{R,L}(r) \sim r^{-\alpha}$ (see Fig.~\ref{fig:m=0.25_dJx=0.1}(c),(d)). Note that the correlator $C_L(r)$ is equal for the legs $j=0$ and $j=1$ due to absence of disorder, and the correlator $C_R(r)$ splits into two branches due to OBC. Both branches of $C_R(r)$ decay faster than $r^{-2}$, thereby indicating that rung disorder is an irrelevant perturbation. This follows from the scaling criterion similar to Harris: for a correlator $C_O(r) \sim r^{-2\Delta}$, disorder is irrelevant if $\Delta > d$, with $d$ being the spatial dimension along which disorder is applied (here $d=1$). On the contrary, the correlator $C_L(r)$ decays significantly more slowly than $r^{-2}$, making the leg disorder relevant.}

To verify this scenario numerically, we introduce disorder only in the leg tunneling amplitudes $J_{xij}$ with a fixed disorder strength $\delta J_x = 0.1$, while keeping the rung tunneling amplitudes $J_{yi} = J_y$ disorder-free. As before, we fix the unit of energy by setting the average tunneling strength $\bar{J}_x = 1$. In Fig.~\ref{fig:m=0.25_dJx=0.1}, we show the behavior of the disorder-averaged entanglement entropy $\bar{\mathcal{S}}$ and the leg magnetization $\bar{S}_L$ for different system sizes. Both quantities exhibit minimal finite-size effects and vary smoothly as functions of $J_y$, indicating that the Ising transition is destroyed by disorder.

\begin{figure}
    \centering
    \includegraphics[width=0.8\linewidth]{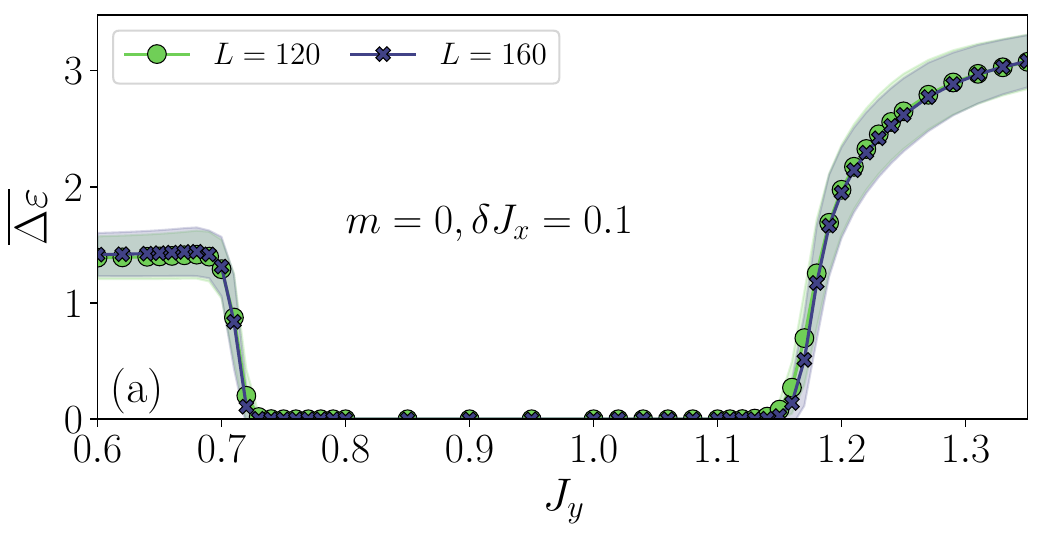}
    \includegraphics[width=0.49\linewidth]{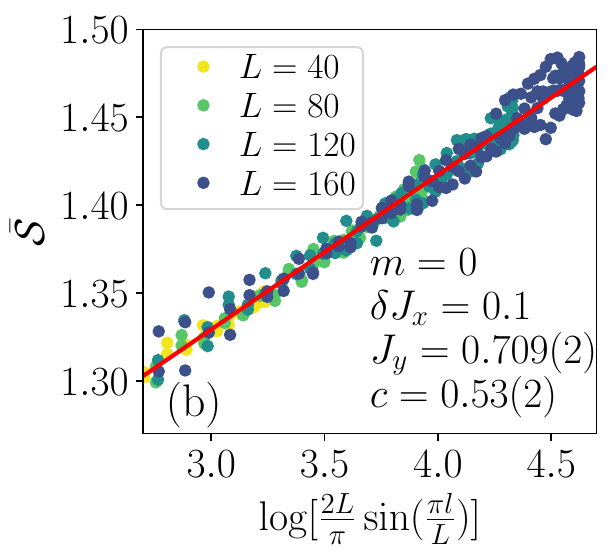}
    \includegraphics[width=0.49\linewidth]{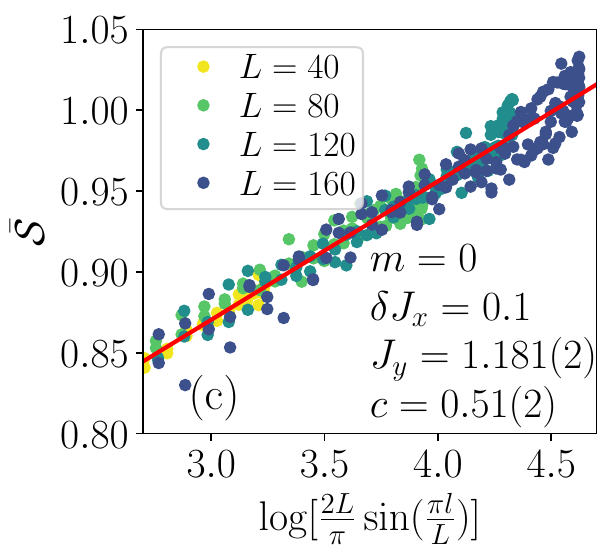}
    \caption{(Color online.) (a) Disorder-averaged entanglement gap $\overline{\Delta \varepsilon}$ at the center of the ladder for system sizes $L=120, 160$ at zero mass in presence of bond disorder on the legs  with disorder strength $\delta J_x=0.1$. Shaded regions denote error bars.
    (b)-(c) The logarithmic scaling of disorder-averaged entanglement entropy $\bar{\mathcal{S}}$ with chord length at VA-to-SPT and SPT-to-V0 transitions for leg disorder strength $\delta J_x=0.1$, respectively. The central charges extract from both the panels are consistent with Ising transition.}
    \label{fig:Egap_m=0_dJx=0.1}
\end{figure}

The situation is drastically different in the massless case. As in the case of disorder on the rungs, disorder in the leg tunneling amplitudes $J_{xij}$ does not spoil the $\mathbb{Z}_2 \times \mathbb{Z}_2$ symmetry that protects the SPT phase. Consequently, for weak disorder strength $\delta J_x$, the SPT phase remains stable, as demonstrated in Fig.~\ref{fig:Egap_m=0_dJx=0.1}(a) through the vanishing of the disorder-averaged entanglement gap $\overline{\Delta \varepsilon}$. Moreover, the critical transitions at the boundaries of the SPT phase also survive. The critical entanglement scaling, shown in Fig.~\ref{fig:Egap_m=0_dJx=0.1}(b) and (c), at these two transition points yields central charges consistent with Ising criticality. Furthermore, finite-size scaling (not shown) of the disorder-averaged magnetizations $\bar{S}_R$ and $\bar{S}_L$ yields a correlation-length critical exponent $\nu \approx 1$ at both transitions, in agreement with the Ising universality class.

As emphasized earlier, in the absence of a microscopic derivation of the field-theoretic description for the ladder system, we do not have a controlled explanation for the stability of the Ising transitions at zero fermion mass that apparently bypasses the Harris bound. Nevertheless, the robustness of the Ising criticality in this regime can be understood phenomenologically from the fact that the protecting $\mathbb{Z}_2 \times \mathbb{Z}_2$ symmetry, and hence the SPT phase itself, remains stable against disorder.

\section{Conclusion}
\label{sec:conclusion}

In this work, we have revisited the quantum link model on a two-leg ladder introduced by Cardarelli et al. \cite{Cardarelli2017, Cardarelli2020}. By a detailed study of the system properties, we determine the possible phases in the system, confirming the existence of a symmetry-protected topological phase for vanishing mass and finding possible transitions in the phase diagram when the mass or leg tunneling is varied. The central charge and correlation-length critical exponent found indicate that the phase transitions observed are of the Ising universality class. Subsequently, we have studied the effect of introducing disorder in the rung tunneling, providing, we believe, the first study of the effect of the disorder on the ground state of the link model. 
It turns out that the SPT order observed for zero mass is robust against a weak disorder, as exemplified by the degenerate entanglement spectrum, which loses this property only for quite strong disorder. Similarly, we have found that the phase transitions between different vortex phases, in the presence of disorder, remain in the Ising class as verified by explicit determination of the central charge. We have also shown that the strong disorder mixes various possible phases in the system for non-zero mass, too. This behavior, in an apparent contradiction with the Harris criterion that would suggest that disorder should destroy the Ising transitions, has been explained by field theoretical arguments.
On the other hand, the leg tunneling disorder seems to strongly affect the physics
for non-zero mass - the criticality disappears this time as expected. Still, for the zero-mass case, SPT phase survives the disorder in leg tunnelings and the transitions to vortex/antivortex states remain within the Ising universality class. Altogether our results demonstrate a subtle interplay between different ways the disorder may be imposed on the ladder system and the remaining symmetries.

\acknowledgments
The work of M.V.R. was funded by the National Science Centre, Poland, under the OPUS call within the WEAVE program 2021/43/I/ST3/01142.

L.T. acknowledges support from the Proyecto Sinérgico CAM Y2020/TCS-6545 NanoQuCo-CM, the CSIC Research Platform on Quantum Technologies PTI-001, and from the Grant TED2021-130552B-C22 funded by MCIN/AEI/10.13039/501100011033 and by the ``European Union NextGenerationEU/PRTR'', and Grant PID2021-127968NB-I00 funded by MCIN/AEI/10.13039/501100011033.

M.L. acknowledges support from: European Research Council AdG NOQIA; MCIN/AEI (PGC2018-0910.13039/501100011033, CEX2019-000910-S/10.13039/501100011033, Plan National FIDEUA PID2019-106901GB-I00, Plan National STAMEENA PID2022-139099NB, I00, project funded by MCIN/AEI/10.13039/501100011033 and by the “European Union NextGenerationEU/PRTR" (PRTR-C17.I1), FPI); QUANTERA DYNAMITE PCI2022-132919, QuantERA II Programme co-funded by European Union’s Horizon 2020 program under Grant Agreement No 101017733); Ministry for Digital Transformation and of Civil Service of the Spanish Government through the QUANTUM ENIA project call - Quantum Spain project, and by the European Union through the Recovery, Transformation and Resilience Plan - NextGenerationEU within the framework of the Digital Spain 2026 Agenda; Fundació Cellex; Fundació Mir-Puig; Generalitat de Catalunya (European Social Fund FEDER and CERCA program, AGAUR Grant No. 2021 SGR 01452, QuantumCAT \ U16-011424, co-funded by ERDF Operational Program of Catalonia 2014-2020); Barcelona Supercomputing Center MareNostrum (FI-2023-3-0024), HORIZON-CL4-2022-QUANTUM-02-SGA PASQuanS2.1, 101113690, EU Horizon 2020 FET-OPEN OPTOlogic, Grant No 899794, EU Horizon Europe Program (grant agreement No 101080086 NeQSTGrant Agreement 101080086 — NeQST); ICFO Internal “QuantumGaudi” project; European Union’s Horizon 2020 program under the Marie Sklodowska-Curie grant agreement No 847648;
“La Caixa” Junior Leaders fellowships, La Caixa” Foundation (ID 100010434): CF/BQ/PR23/11980043.

The work of J.Z. was funded by the National Science Centre, Poland, project 2021/03/Y/ST2/00186 within the QuantERA II Programme that has received funding from the European Union Horizon 2020 research and innovation programme under Grant agreement No 101017733.  A support by the Strategic Programme Excellence Initiative (DIGIWorkd) at Jagiellonian University is acknowledged.

T.C. acknowledges the financial support from the Young Faculty Initiation Grant (NFIG) at IIT Madras (Project No. RF24250775PHNFIG009162) and the Anusandhan National Research Foundation (ANRF), India via the Prime Minister Early Career Research Grant Scheme ANRF/ECRG/2024/001198/PMS.

We gratefully acknowledge Polish high-performance computing infrastructure PLGrid (HPC Center: ACK Cyfronet AGH) for providing computer facilities and support within the computational grant no. PLG/2025/018400.
The numerical simulations have been performed using the TeNPy library \cite{tenpy}.

\section{DATA AVAILABILITY}
The data presented in this work are available at Ref. \cite{data}
\appendix

\section{Finite-size scaling analysis}
\label{app:fss}

\tc{For the finite-size scaling analysis performed in this work, we follow the approach outlined in Refs.~\cite{Suntajs2020, Sai2021}, which is applicable when the observable under consideration is monotonic across the phase transition. Following the standard finite-size scaling hypothesis, for a given observable $Q$, we have
\begin{align}
    Q L^{\beta/\nu} \equiv Q^{\text{scaled}} = f\bigl((J_y - J_y^c)\, L^{1/\nu}\bigr),
\end{align}
where $f(.)$ is a scaling function. If $Q^{\text{scaled}}$ is a smooth, continuous, and monotonic function of $(J_y - J_y^c)\, L^{1/\nu}$, then the sequence of rescaled data points $\{Q^{\text{scaled}}_j\}$, ordered according to nondecreasing values of $(J_y - J_y^c)\, L^{1/\nu}$, satisfies
\begin{align}
    \sum_j \left| Q^{\text{scaled}}_{j+1} - Q^{\text{scaled}}_j \right|
    = \max_j \{Q^{\text{scaled}}_j\} - \min_j \{Q^{\text{scaled}}_j\}.
\end{align}
Deviations from this equality provide a quantitative measure of the quality of the data collapse. Accordingly, we define the cost function for the finite-size scaling analysis as
\begin{align} 
    \mathcal{C}(J_y^c, \beta, \nu) = \Big[ & \sum_j \left| Q^{\text{scaled}}_{j+1} - Q^{\text{scaled}}_j \right| \nonumber \\ 
    & - \left( \max_j \{Q^{\text{scaled}}_j\} - \min_j \{Q^{\text{scaled}}_j\} \right) \Big]^2. 
\end{align}
The optimal estimates of the critical point $J_y^c$ and the critical exponents $(\beta,\nu)$ are obtained by minimizing this cost function. Since $\mathcal{C}(J_y^c, \beta, \nu)$ may exhibit multiple local minima, we first perform a coarse-grained brute-force global search over a predefined parameter range, and subsequently refine it using a local optimization algorithm, such as the Nelder--Mead simplex method, to obtain the final estimates.
}

\tc{To obtain reliable error bars for the estimated parameters, we adopt a statistical bootstrapping procedure. In the disorder-free case, for each system size $L$, we generate bootstrap samples by resampling the available data points in $J_y$ (with replacement), thereby constructing new datasets $Q(L, J_y)$. For each such resampled dataset, we repeat the full minimization of the cost function to obtain estimates of the parameters. This procedure is repeated over $100$ bootstrap iterations, and the standard deviations of the resulting parameter distributions are taken as the corresponding error bars.}

\tc{In the presence of disorder, we additionally incorporate the statistical uncertainties arising from disorder averaging. For each $(L,J_y)$, we perturb the observables by adding Gaussian noise with standard deviations equal to the estimated errors from disorder averaging during each resampling step. The bootstrap analysis is then repeated over $500$ iterations to ensure convergence of the estimated uncertainties.}

%

\end{document}